\documentclass[useAMS,usegraphicx,usenatbib]{mn2e}

\usepackage{times,subfigure}
\usepackage{threeparttable}
\usepackage{epsfig}
\usepackage{textcomp}
\usepackage{amssymb}
\usepackage{wrapfig}

\usepackage{rotate}
\usepackage{lscape}
\usepackage{rotating}
\usepackage{color}
\usepackage{longtable}
\usepackage{amsmath}

\newif\ifAMStwofonts

\AMStwofontstrue

\def\deg{$^{\circ}$}

\def\xmm{{\it XMM-Newton}}
\def\suz{{\it Suzaku}}

\newcommand{\dc}{\Delta \chi^{2}}
\newcommand{\rg}{r_{\mathrm{g}}}

\newcommand{\rin}{r_{\rm in}}
\newcommand{\rout}{r_{\rm out}}
\newcommand{\rbr}{r_{\rm br}}

\newcommand{\de}{^{\circ}}

\title[On the absence of broad lines in Seyferts]{On the apparent absence of broad iron lines in Seyfert galaxies}

\author[S. Bhayani et~al.]{S. Bhayani$^1\thanks{shyam.bhayani03@imperial.ac.uk}$ and K. Nandra$^2$ \\
$^1$Astrophysics Group, Imperial College London, Blackett Laboratory, Prince Consort Road, London, SW7~2AW, UK \\ 
$^2$Max-Planck-Institut f\"{u}r extraterrestrische Physik, Giessenbachstrasse 1, D-85748, Garching bei M\"{u}nchen, Germany\\}

\date{}

\pubyear{2011}

\begin{document}

\maketitle

\label{firstpage}

\begin{abstract}

We present an analysis of $\xmm$ observations of eleven Seyfert galaxies that appear to be missing a broad iron K$\alpha$ line. These objects represent a challenge to the established paradigm for active galactic nuclei, where a relatively cold accretion disc feeds the central black hole. In that paradigm, X-ray illumination of the accretion disc should lead to continuum and fluorescence emission from iron which is broadened and shifted by relativistic effects close the hole. We extend the work of  \citet{na07}, who found no evidence for such a component in an earlier analysis of these objects, by testing a variety of more complex relativistic reflection models. Specifically, we consider the possibility that the disc is highly ionised, and/or that the the reflection is heavily blurred by strong relativistic effects in a Kerr geometry.  We find that in 8/11 of the observations with no apparent broad iron line, the fit is significantly improved when an ionised or strongly blurred reflector is included, and that all 11 observations allow for such a component. The disc inclinations are found generally to be around $60\de$, which when combined with a steep emissivity profile results in strong relativistic blurring of the reflection, rendering the K$\alpha$ line difficult to distinguish from the underlying continuum. Nevertheless, relativistic reflection does appear to be present, and the strength of the smeared reflection is similar to that expected from a flat disc illuminated by a point source. Such blurred reflection and the associated steep radial emissivity profiles are consistent with the gravitational light bending of the continuum photons close to the black hole. 
\end{abstract}


\section[Introduction]
{Introduction}

A key feature in the X-ray spectra of Active Galactic Nuclei (AGN) is a Compton reflection component with a relativistically broadened iron line. The line arises via illumination of the accretion disc by the X-ray continuum \citep{fa89, la91} and can provide important insights into the inner regions of AGN, as it is broadened and shaped by Doppler effects and the strong gravitational field of the central black hole. The first clear example was seen in the {\it ASCA} spectrum of MCG-6-30-15 \citep{ta95}, and subsequent analysis of an AGN sample observed by {\it ASCA} suggested that this phenomenon was common \citep{na97_1}. On the other hand, the {\it ASCA} spectra already hinted at a diversity in the iron line profiles. For example, the spectra were consistent with a range of radial emissivity indices, which is commonly used to parameterize how close to the central black hole the line flux is generated.

Subsequent observations with better statistics and/or higher spectral resolution have been provided by \textit{XMM-Newton}, \textit{Chandra} and \textit{Suzaku}. These have revealed broad and skewed iron lines in many individual objects \citep[e.g.][]{fa02, mi07}. On the other hand, they have also shown that broad iron lines of this kind are not always found. For example, the X-ray spectra in certain observations of NGC 7213 \citep{bi08}, NGC 3227 \citep{ma09} and NGC 5548 \citep{po03, li10} seem to possess only a narrow iron line. This line component is thought to originate in a distant reflector such as the molecular torus \citep{kr87} while the broad and relativistic iron line appears to be missing.

A systematic approach to this problem was provided via an \textit{XMM-Newton} survey of local Seyfert 1 galaxies by \citet[][hereafter N07]{na07}. This work showed that $\sim$ 70\% of objects possessed a broadened iron line of some kind. However, out of these, in only around 70\% could the broadening be classified as relativistic. Around a quarter of the sample showed an iron line that, while significantly broadened, was consistent with most of the emission originating beyond 50$\rg$ from the black hole. A clear question then arises of why there is apparently no line emitting gas within $50\rg$, where most of the accretion power -- and X-ray emission -- is generated. Perhaps most troublingly, a similar proportion ($\sim 30\%$) of the sample showed no evidence for broadened iron line emission at all, with only the narrow iron line from distant reflection being detectable.

The results for MCG-6-30-15 and other broad line sources \citep{ca10} have provided strong support for the standard paradigm for AGN, where an accretion disc feeds a central supermassive black hole, and indeed even for the existence of black holes themselves. Conversely, sources which lack such features challenge that paradigm. The presence of a broad iron line is an almost inevitable consequence of the basic physical processes thought to operate in the central regions, and the absence of this feature makes it very difficult to fit all objects into a single unifying framework. 

N07 suggested that the apparent lack of a disc line in some cases may be due to them being viewed at a high inclination. This is expected to lead to weaker reflection. In combination with low S/N spectra, it may then be difficult to detect the reflection component in these observations. The latter conclusion has been strongly supported by \citet{ca10}, who clearly demonstrated that very high signal-to-noise ratio spectra are needed to detect the broad emission. Alternatively, a relativistically broadened iron line would not be able to form if the accretion disc is very highly ionised \citep{ro93, zy94}. This could occur if the X-ray emission originates in magnetic flares close to the surface of the disc.

It is also possible that the iron line in the ``narrow-line" observations is broadened to such an extent that it is difficult to distinguish from the continuum. This can occur if the accretion disc is viewed at a high inclination. Similarly, an iron line will be heavily smeared if there is a strong concentration of X-ray emission in the central regions. For example, this is possible in a Kerr geometry via light bending \citep{mi04}. In this case, the relativistic signature is not missing but is further evidence for the extreme gravitational effects at the centre of AGN.

The cases of high ionisation and extreme smearing of the line in a Kerr geometry were not considered by N07, who limited themselves to the simpler cases of a neutral disc, and a ``lamppost" emissivity in which the emission is spread over a wide range of disc radii. The aim of this paper is to extend the analysis of N07 by exploring these possibilities.

\section[Observations]
{Observations}

The current study comprises observations of the 11 objects from the N07 sample which showed no evidence for broad iron line emission. The reduction methods are fully described in their work, but briefly, we only include local Seyferts $\left( z < 0.05 \right)$ with observations available in the \xmm\ public as of 1 Jan 2006. Seyfert 2s are excluded due to the obscuration of the inner regions by the dusty torus. The observations were recorded with the EPIC-pn CCD camera and details such as the data, exposure, mode and filter are given in Table 2 of N07. After screening for periods of high background, an observation was rejected if it did not possess a minimum of 30,000 counts between 2 -- 10~keV. The analysis is restricted to 2.5 -- 10~keV as the focus is on the iron K$\alpha$ line and associated reflection component.

\section[Models and Results]
{Models and Results}

\subsection{Model NBR}

For the X-ray spectral fitting, we used XSPEC v11 \citep{ar85}. The models that we tested are compared to a model that lacks a broad iron line and its associated reflection component (Model NBR - ``No Blurred Reflection"). In XSPEC terminology, Model NBR is described as:

\begin{align}
\label{eq:xspec_nbr}
\nonumber \mbox{Model NBR} =\ & \mathrm{WABS} \times (\mathrm{POWERLAW + PEXMON})\\
& \times \mathrm{CWA18}.
\end{align}

\noindent The X-ray continuum is represented by the ``POWERLAW" component with WABS accounting for the neutral absorption. The Galactic $N_{\rm H}$ values are given in Column 5 of Table 1 in N07 and taken from \citet{el89}, \citet{st92} and \citet{di90}. 4/11 of the narrow-line observations also required ionised absorption and this is modelled by the table model ``CWA18", which was developed by N07 using the photoionisation code XSTAR \citep[][see N07 for further details]{ka04}. The ``PEXMON" component accounts for the continuum reflection from the distant material, which was found by N07 to be ubiquitous in their sample. This is a model that N07 developed by combining the XSPEC ``GAUSSIAN" and ``PEXRAV" \citep{ma95} models. The advantage of using PEXMON is that it ties the strength of the iron line and Compton reflection together, providing a self-consistent model of the reflection continuum and  the iron line (see N07 for further details.). Solar abundances were assumed along with a high-energy cut-off to the continuum of $E_{\rm Cut}$ = 1~MeV and an inclination of $i = 60\de$. The only free parameter of PEXMON is the reflection strength, $R$. Narrow ($\sigma = 10$~eV) emission or absorption lines were added by N07 for some observations if they made an improvement to the fit. The energies of these lines are given in Table 7 of N07. The model components and parameters of Model NBR and the subsequent models are listed in Table \ref{tab:mods}.

\begin{figure*}

\caption{The unfolded 2.5 -- 10~keV spectrum and data/model ratio for the 8/11 observations that are well-fit with strongly smeared reflection and/or an ionised disc. The spectrum of Model ION is shown if Model SRI did not give an acceptable fit i.e. NGC 7213. The power-law continuum is shown with a dotted line. The blurred and distant reflection are shown with red dashed and dash-dotted lines respectively. The total spectrum is shown as a solid line.}

\includegraphics[width=80mm]{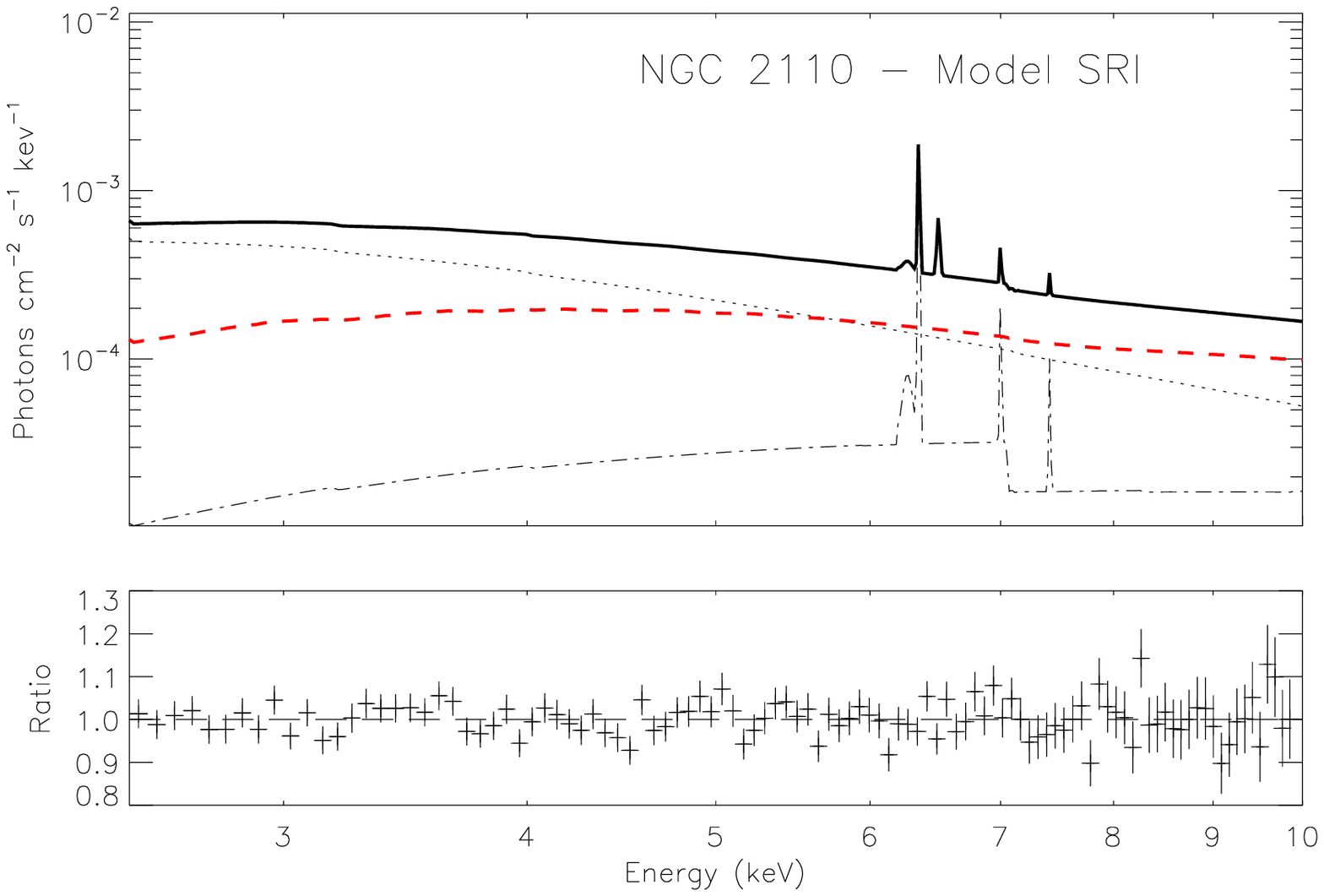}
\includegraphics[width=80mm]{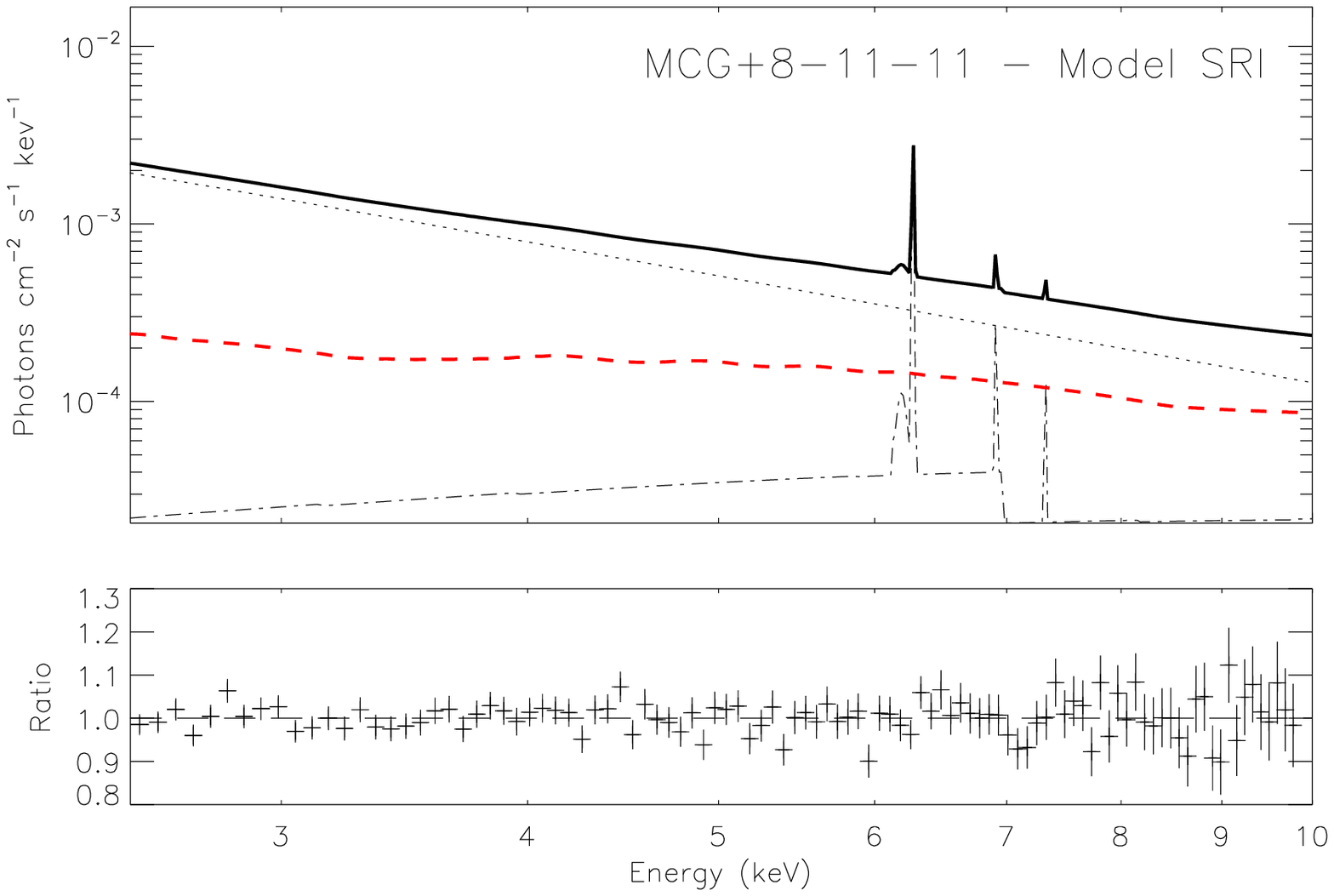}
\includegraphics[width=80mm]{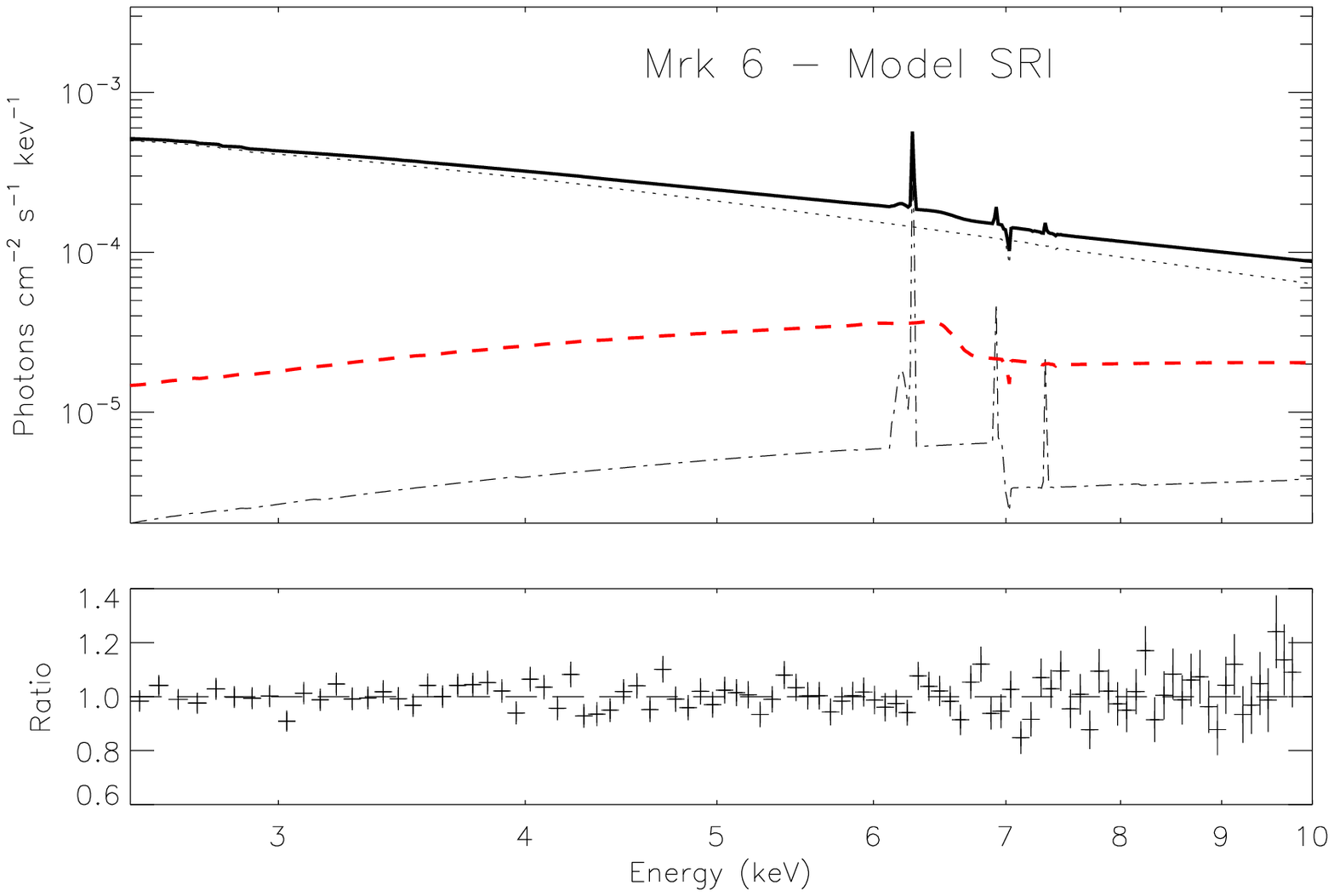}
\includegraphics[width=80mm]{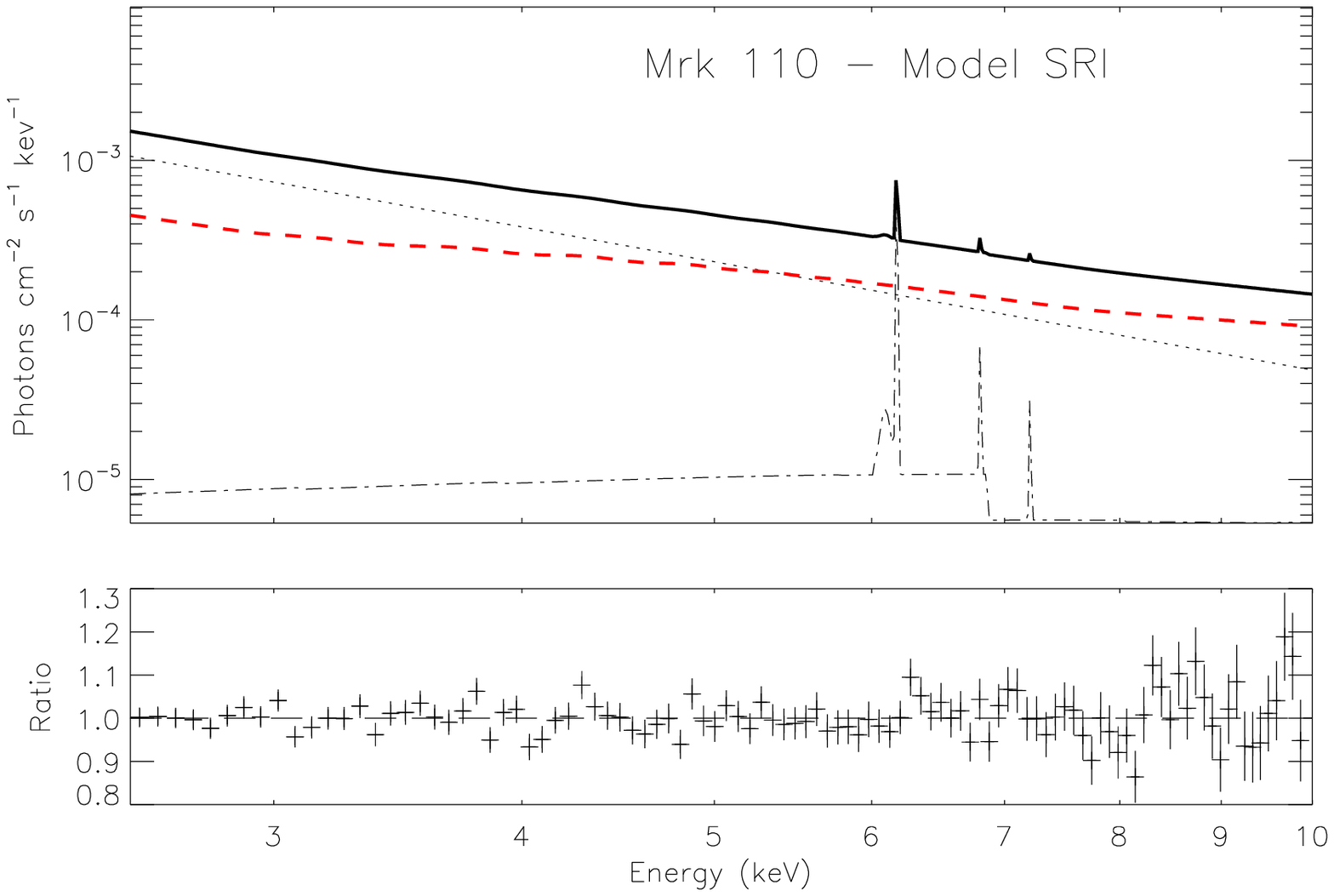}
\includegraphics[width=80mm]{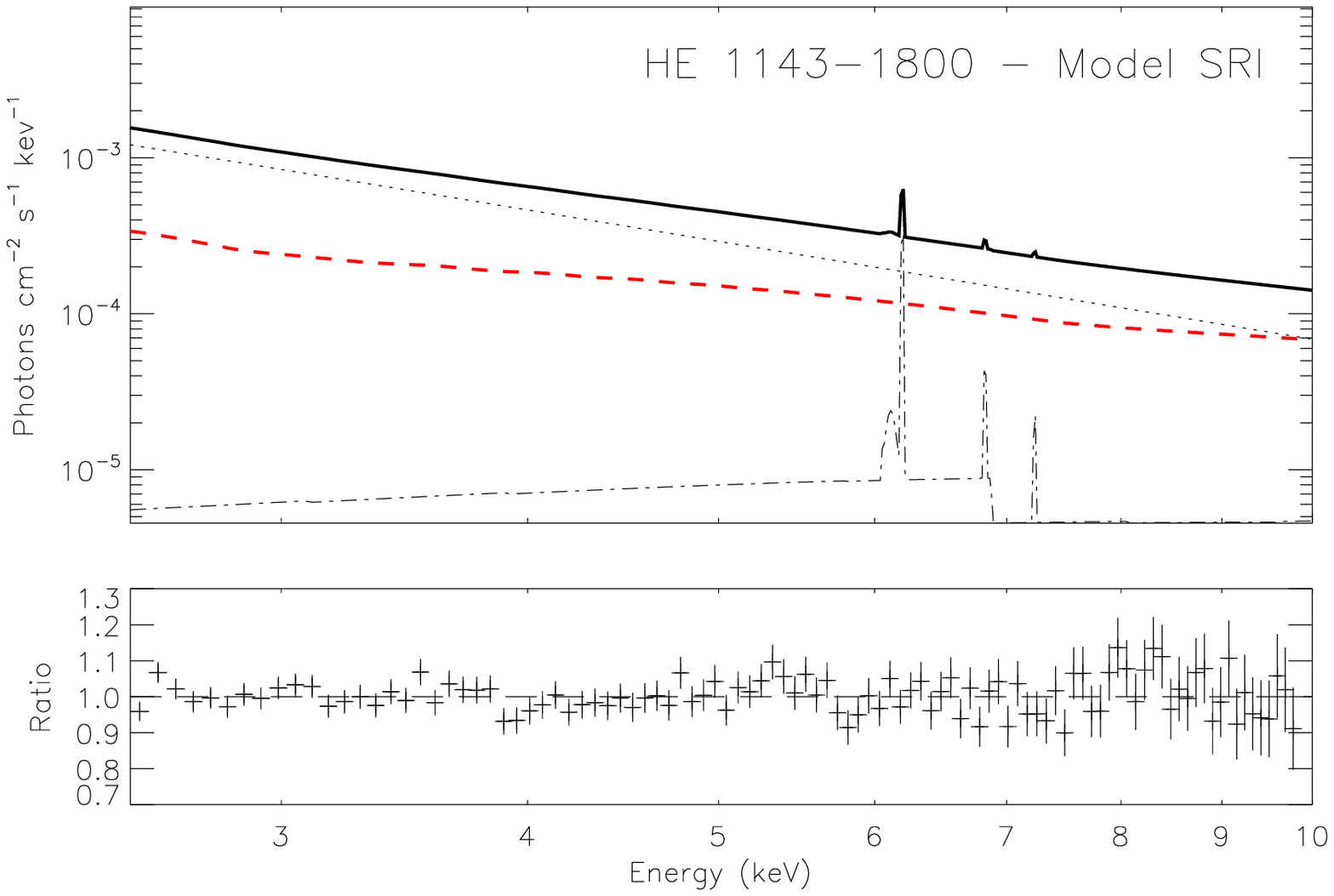}
\includegraphics[width=80mm]{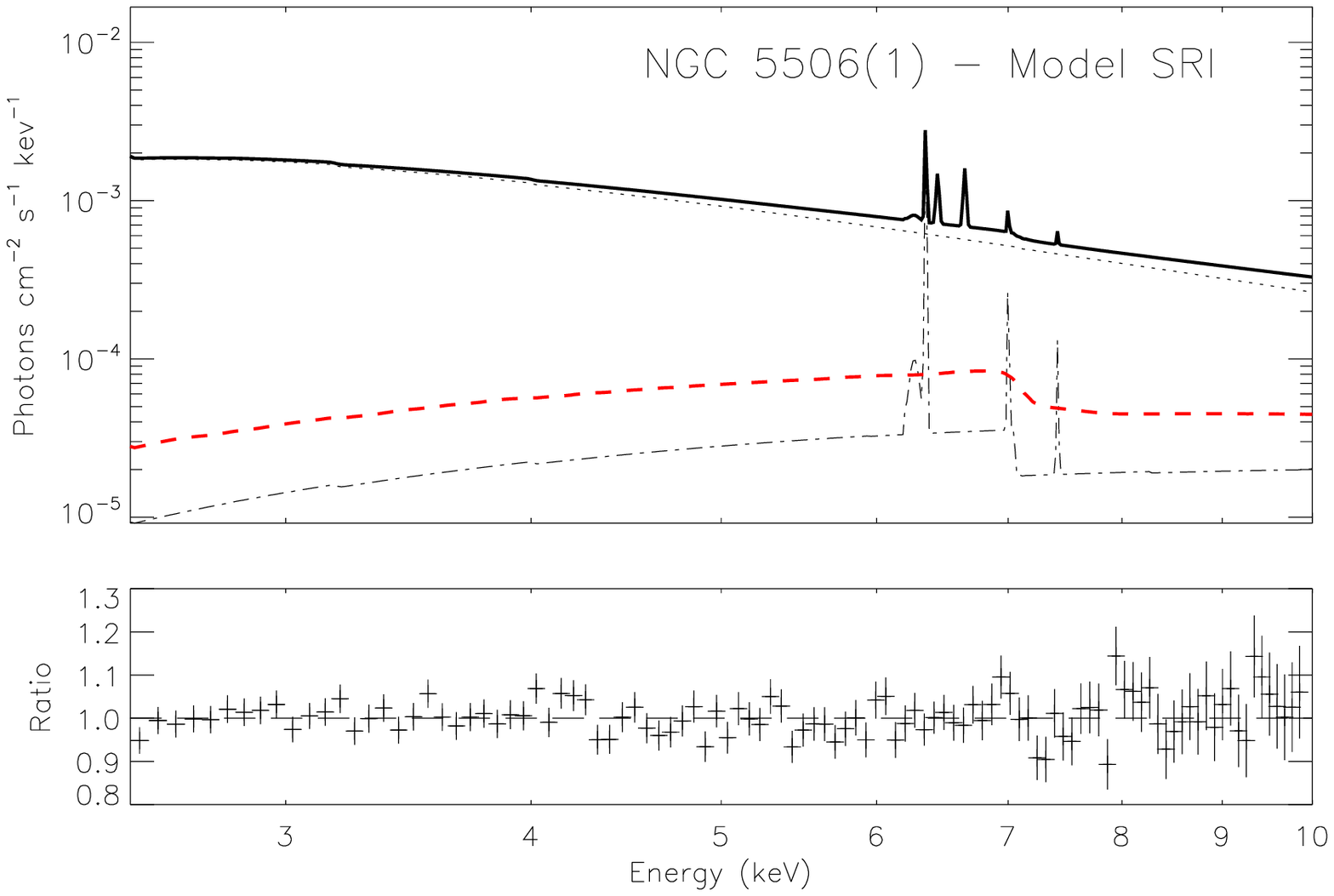}
\includegraphics[width=80mm]{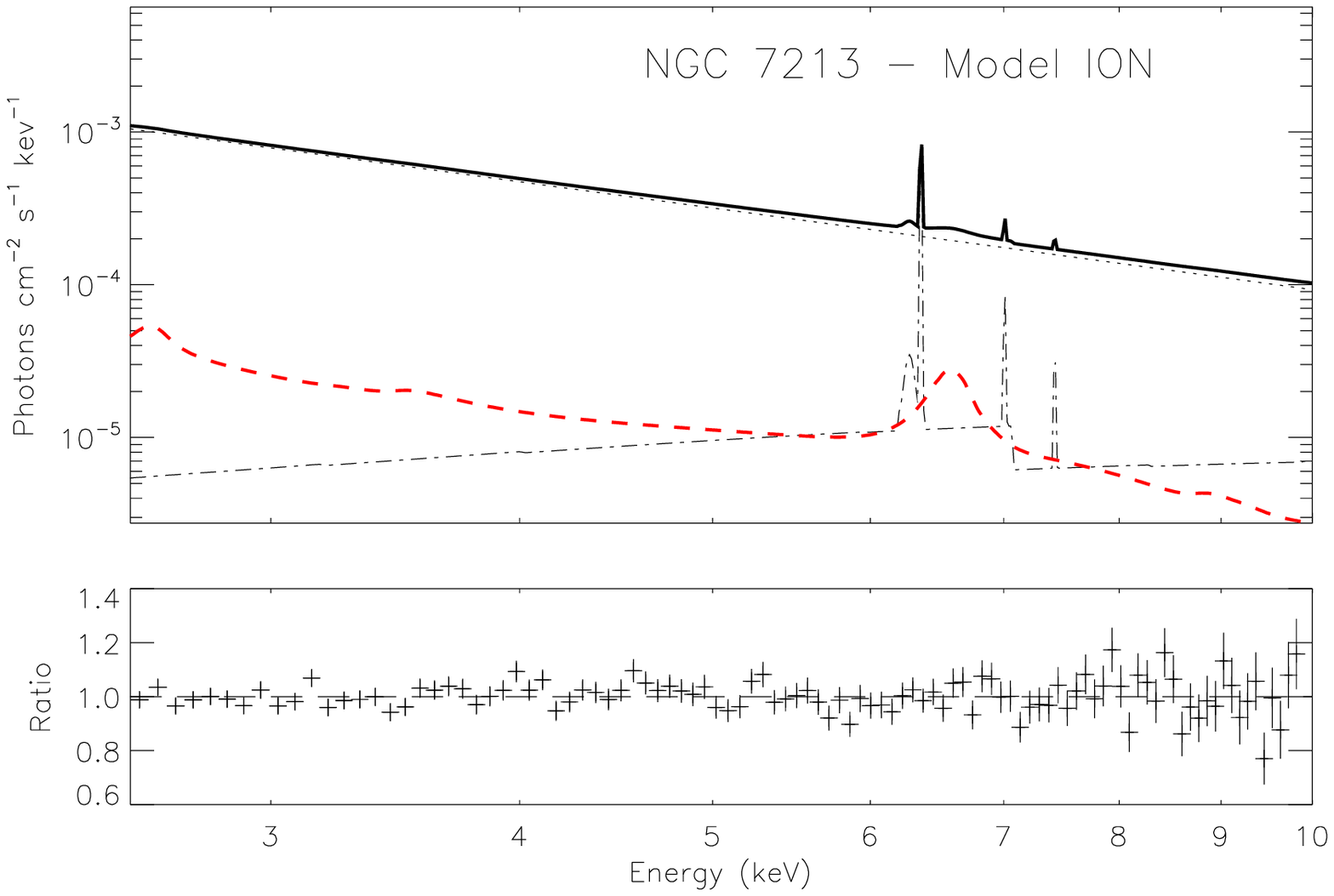}
\includegraphics[width=80mm]{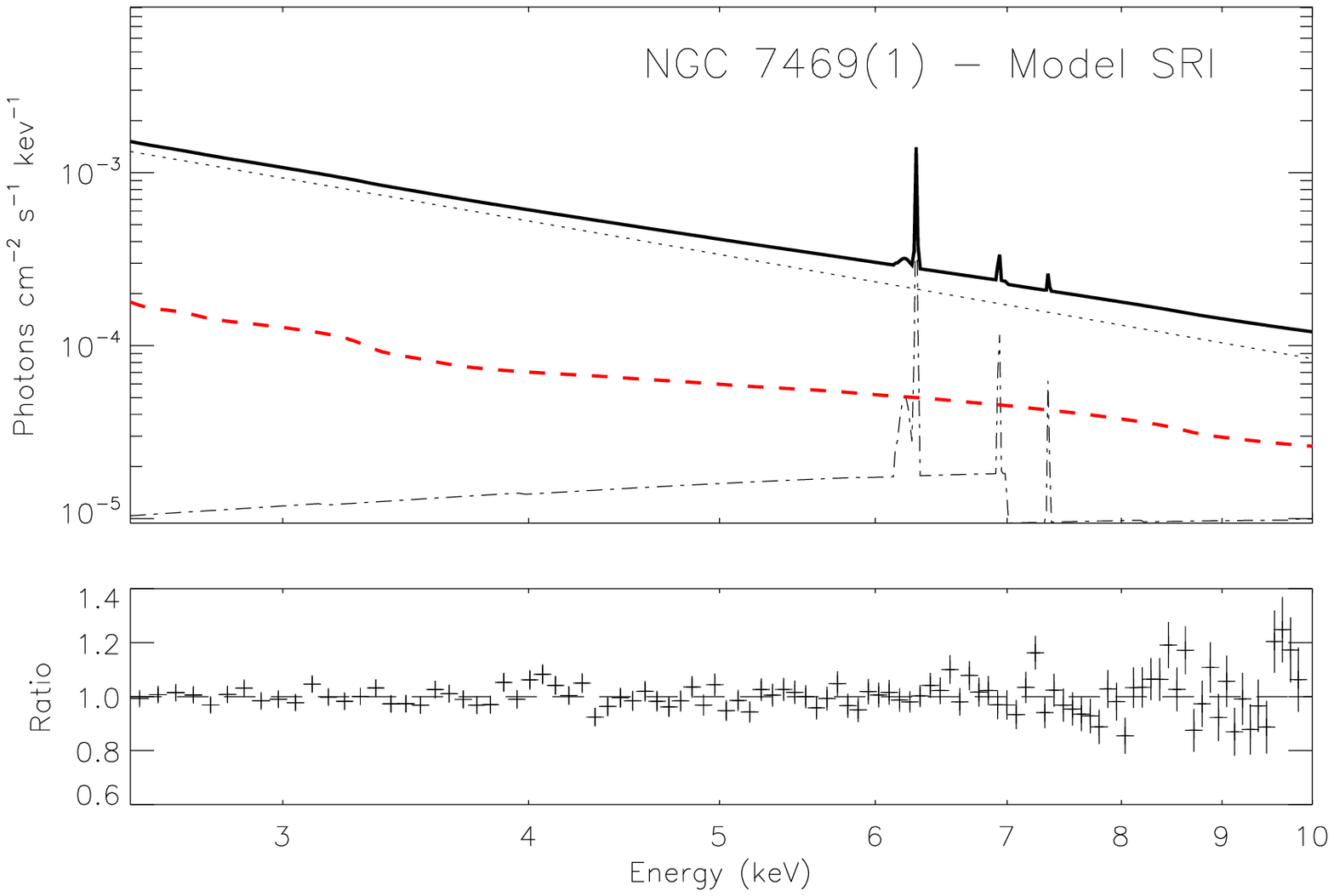}
\label{fig:mo_res1}
\end{figure*}

\begin{figure*}

\caption{The unfolded 2.5 -- 10~keV spectrum and data/model ratio for the 3/11 observations that do not show a significant improvement when fit with strongly smeared reflection and/or an ionised disc. The spectrum of Model SRI is shown with the identity of the model components being the same as in Figure \ref{fig:mo_res1}. The blue long dashed-line shows the 90\% upper limit on the REFLIONX model component.}

\includegraphics[width=80mm]{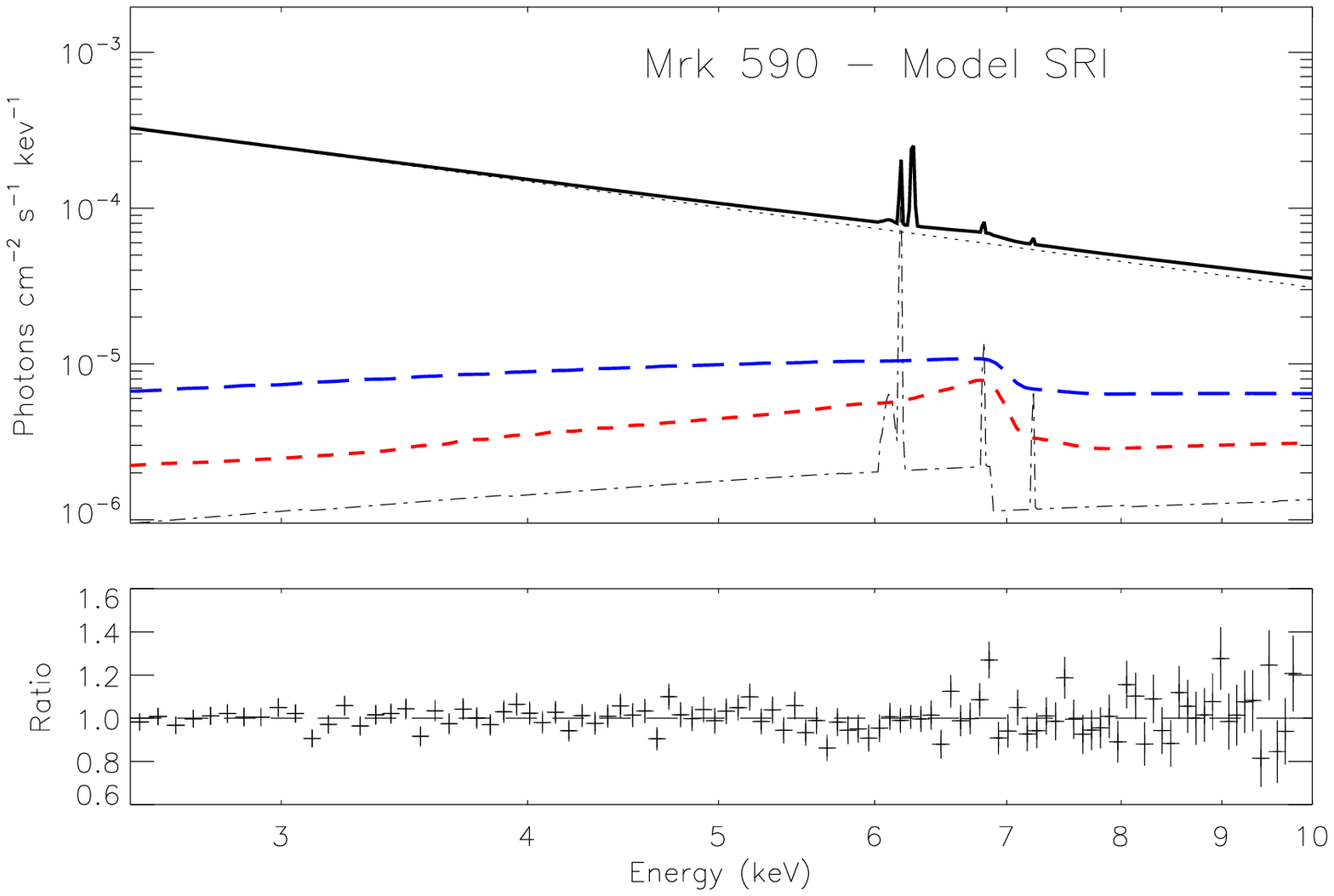}
\includegraphics[width=80mm]{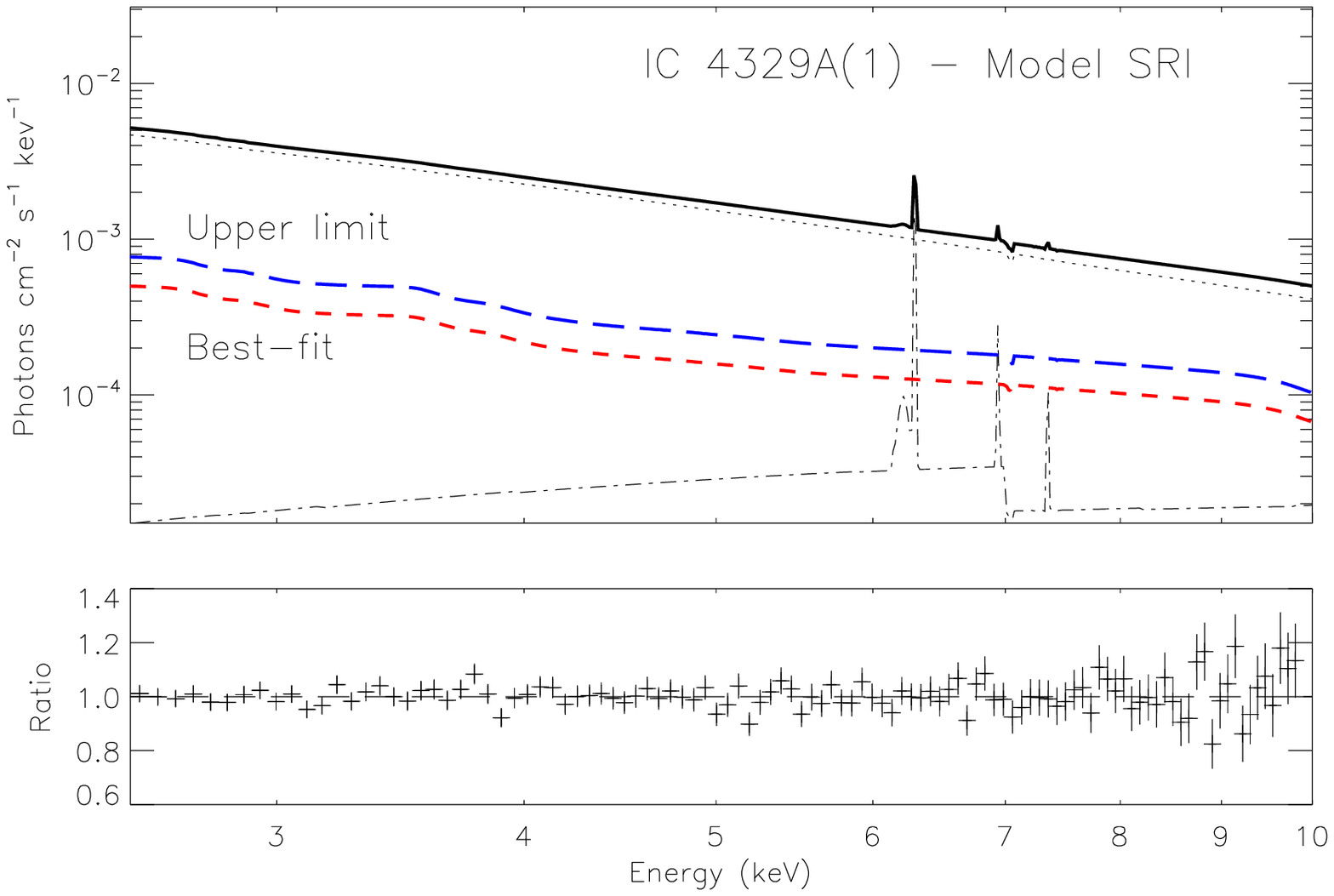}
\includegraphics[width=80mm]{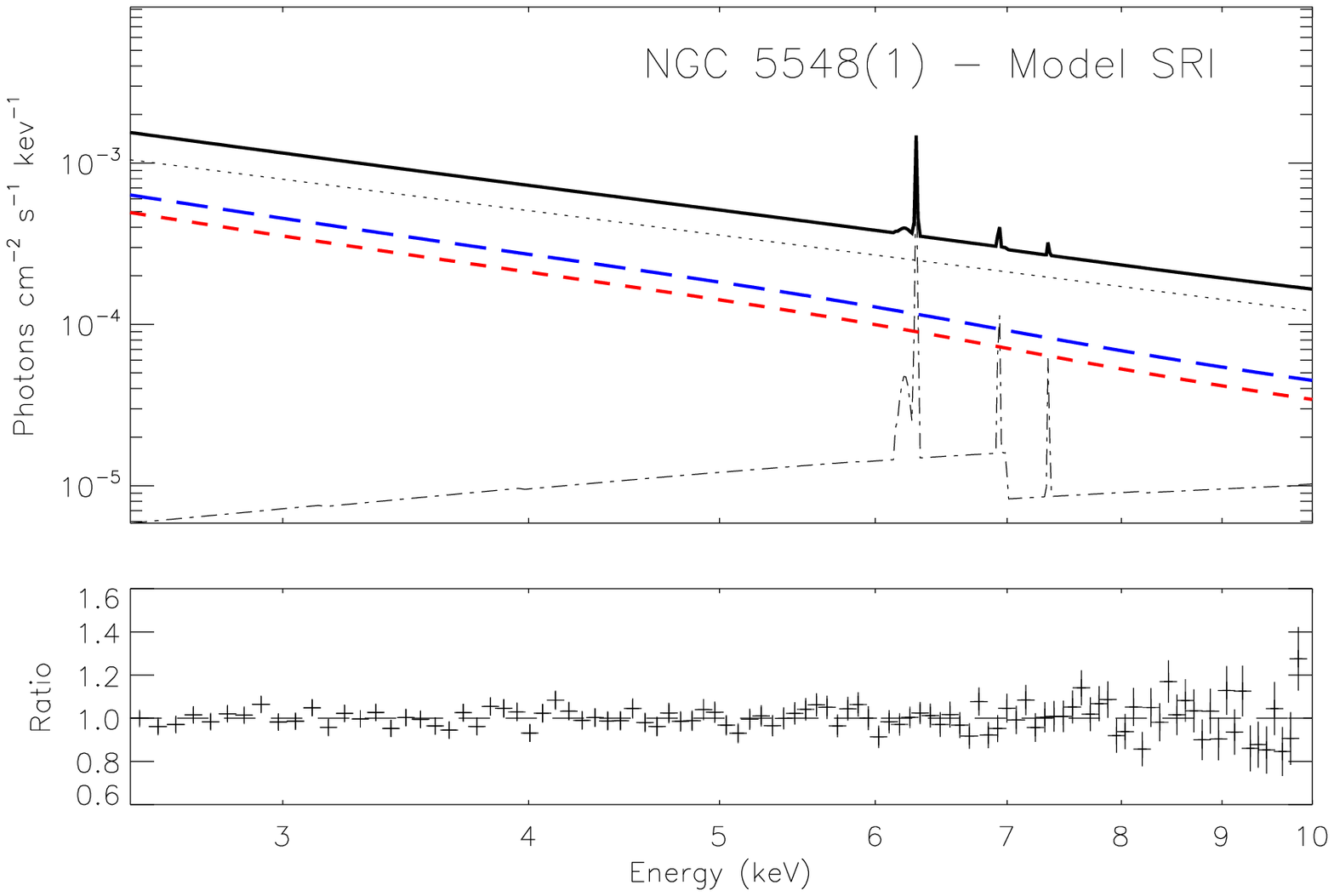}

\label{fig:mo_res2}
\end{figure*}

The best-fit $\chi^{2}$/d.o.f.\ of Model NBR to the 2.5 -- 10~keV energy spectra are shown in Column 1 of Table \ref{tab:re}. The fits of all the models that we tested are compared to Model NBR so we list the $\dc$ of the models relative to Model NBR in columns 3 -- 6. The best-fit disc inclination, reflection strength and ionisation parameter are given in Table \ref{tab:par}. The unfolded spectra of the best-fitting models to the eleven narrow-line observations are shown in Figures \ref{fig:mo_res1} and \ref{fig:mo_res2}.


Errors are calculated at a confidence level of 90\% for one interesting parameter. The significance of the improvement in the fits over Model NBR was assessed using Monte Carlo simulations, which is superior to the F-test in many cases \citep{pr02}, and we followed the approach of N07. Using \textit{fakeit} in XSPEC, we created 10,000 synthetic spectra, based on Model NBR with parameters set to the mean values found by N07 for their whole sample. A spectrum with the same number of counts as the mean of the narrow-line observations ($\sim 70,000$ counts) between 2.5 -- 10~keV was adopted and Poisson noise was added. These simulated spectra were refit with Model NBR and the change in $\chi^{2}$ after fitting with the model of interest was recorded. The $\dc$ value at which 95 per cent of the simulations have a smaller value was set as the threshold $\dc$ for a significant detection of an iron line at 95\% confidence level. Observations with a significant detection are shown in bold in Table \ref{tab:re}.

\begin{table*}

\caption{The XSPEC model components and parameters for each model. Part (a) of the table lists the components with parameters that are the same in all five models and part (b) gives the model components that differ between the models. X denotes parameters that are dependent on the AGN and \textbf{f} represents free parameters. --- is used when a model component is not included and ``red'' denotes a redundant parameter i.e. the break radius for KDBLUR2 when $q_{1} = q_{2}$.}

\label{tab:mods}
\centering
\begin{center}
\begin{tabular}{lcccccc}
\hline

\multicolumn{7}{c}{(a)}\\
Model component & Parameter &  \multicolumn{5}{c}{All models}\\
\hline
WABS & $N_{\rm H}$ & \multicolumn{5}{c}{X} \\
\hline
POWERLAW & $\Gamma$ & \multicolumn{5}{c}{\textbf{f}}  \\
 & Normalisation & \multicolumn{5}{c}{\textbf{f}}\\
\hline
PEXMON (torus) & $E_{\rm Cut}$ (keV) & \multicolumn{5}{c}{1000}\\
 & $z$ & \multicolumn{5}{c}{X}\\
 & Abundance & \multicolumn{5}{c}{1.0} \\
& $i$ ($\de$)& \multicolumn{5}{c}{60} \\
 & Normalisation & \multicolumn{5}{c}{\textbf{f}}\\
\hline
CWA18 (if necessary) & log($N_{\rm H}$) &  \multicolumn{5}{c}{\textbf{f}} \\
 & log($\xi$) &  \multicolumn{5}{c}{\textbf{f}} \\
\hline
\\
\multicolumn{7}{c}{(b)}\\

Model component & Parameter &  \multicolumn{5}{c}{Model}\\
& & NBR & N07 & ION & SRC & SRI \\
\hline
PEXMON (disc) & $E_{\rm Cut}$ (keV) & --- & 1000 & --- & 1000 & --- \\
 & $z$ & --- & X & --- & X & --- \\
 & Abundance & --- & 1.0 & --- & 1.0 & --- \\
 & Normalisation & --- & \textbf{f} & --- & \textbf{f} & --- \\
\hline
REFLIONX  & Abundance & --- & --- & 1.0 & --- & 1.0 \\
 & $\xi$ & --- & --- & \textbf{f} & --- & \textbf{f} \\
 & $z$ & --- & --- & X & --- & X \\
 & Normalisation & --- & --- & \textbf{f} & --- & \textbf{f} \\
\hline
KDBLUR2 & $q_{1}$ & --- & 0 & 0 & \textbf{f} & \textbf{f} \\
 & $q_{2}$ & --- & 3 & 3 &  = $q_{1}$ & = $q_{1}$ \\
& $i$ & --- & \textbf{f} & \textbf{f} & \textbf{f} & \textbf{f} \\
& $\rin$ ($\rg$) & --- & 6 & 6 & 1.24 & 1.24 \\
& $\rout$ ($\rg$) & --- & 400 & 400 & 400 & 400 \\
& $\rbr$ ($\rg$) & --- & \textbf{f} & \textbf{f} & red & red \\
\hline

\end{tabular}
\end{center}

\end{table*}

\subsection{Model N07}

In addition to the model components in Model NBR, the best-fitting model in N07 (either their Model E or F) includes reflection from the accretion disc. This is defined as Model N07 and its XSPEC description is:

\begin{align}
\label{eq:xspec_n07}
\nonumber \mbox{Model N07} =\ & \mathrm{WABS} \times (\mathrm{POWERLAW + PEXMON}\\
&+ \mathrm{KDBLUR2} \times \mathrm{PEXMON) \times CWA18}.
\end{align}

\noindent N07 investigated cold disc reflection, which as with the distant reflection is modelled by PEXMON. The only difference is that the disc inclination is left as a free parameter. The smearing of the iron line and reflection continuum is implemented by the ``KDBLUR2" model component \citep{la91}. N07 considered a lamppost geometry for the accretion disc and corona. This can be approximated by a power-law radial emissivity profile ($\propto r^{-q}$) that breaks from a slope of $q_{1}=0$ to $q_{2}=3$ at the break radius, $\rbr$. A spinning black hole was not found to generally produce a better fit to the energy spectra than a static black hole, so Model N07 set the inner radius of the accretion disc to $\rin = 6\rg$ while the outer radius was fixed to $\rout = 400\rg$.

\subsection{Ionised Disc -- Model ION}

As Model N07 tested only neutral accretion disc reflection, our first extension to the N07 analysis was to determine if an ionised accretion disc model can reveal the reflection in the apparent narrow-line spectra. This model is termed Model ION and differs from Model N07 only in the replacement of the cold disc model, PEXMON, with the ionised reflection model, ``REFLIONX'' \citep{ro05} for the blurred reflection. In REFLIONX, the ionisation state of the disc is parametrised by the ionisation parameter, defined as $\xi = L/nr^{2}$, where $L$ is the luminosity of the ionising continuum, $n$ is the hydrogen number density and $r$ is the distance to the X-ray source. The XSPEC description of Model ION is:

\begin{align}
\label{eq:xspec_ION}
\nonumber \mbox{Model ION} =\ & \mathrm{WABS} \times (\mathrm{POWERLAW + PEXMON}\\
&+ \mathrm{KDBLUR2} \times \mathrm{REFLIONX) \times CWA18}.
\end{align}

Evidence for a broad and ionised reflection component at 95\% confidence is found in five observations: MCG+8-11-11, Mrk 110, NGC 7469(1), Mrk 6 and NGC 7213, with the first three being significant even at 99\% confidence. Excluding NGC 7213, the remaining four of these observations display an iron line that is relativistically broadened, as their characteristic break radii are less than 50$\rg$. The REFLIONX ionisation parameter of these four observations is mild, ranging from $\xi= 12-200$~erg cm s$^{-1}$. Nonetheless, accounting for the ionisation state of the disc may be very important, as even in a mildly ionised disc, Compton scattering can contribute to additional broadening \citep{ro99}. The REFLIONX reflection strengths of the accretion disc in MCG+8-11-11, Mrk 6, Mrk 110 and NGC 7469(1) are between $R = 0.55-0.97$, which is slightly less than the strength, $R = 1$, expected by a flat accretion disc subtending a solid angle of 2$\pi$ below a central point source. They are however, typical of reflection fractions for the objects with visible and detectable neutral broad lines in the N07 sample. NGC 7213 differs from the well-fit observations in requiring a smaller reflection strength of $R = 0.11$.

The disc inclination is well determined in the case of Mrk 110 at \textit{i} = 36\deg$^{+19^{\circ}}_{-12^{\circ}}$. MCG+8-11-11 and Mrk 6 are much less well determined, but the best-fit values are $45\de$ and $26\de$ respectively. These are close to the average disc inclination for the N07 broad line sources. Hence the spectra of MCG+8-11-11, Mrk 6 and Mrk 110 show properties entirely typical of sources with broad, relativistic reflection components, with the sole difference being that the reflector is ionised. The inclinations of the discs in NGC 7213 and NGC 7469(1) peg at the model limits $i = 0\de$ and 85$\de$ respectively and thus the detections in these cases cannot be considered robust. If the disc inclination is high, the reflection would be expected to be very weak and very broad. An alternative possibility, that we examine in the following section, is that the extreme broadening is due to stronger relativistic effects than assumed by the models in N07.

\begin{table}

\caption{The best-fit $\chi^{2}$/d.o.f.\ of Model NBR and the change ($\Delta \chi^{2}$/d.o.f.) in goodness-of-fit for Models N07, ION, SRC and SRI relative to Model NBR. Significant improvements at a confidence level of 95\% are shown in bold.}

\label{tab:re}
\centering
\begin{center}
\begin{tabular}{lccccc}
\hline

Name & \multicolumn{5}{c}{Model}\\
 & NBR & N07 & ION & SRC & SRI \\
\hline
Mrk 590 & 95.9/98 & 3.2/3 & 3.2/4  & 1.8/3 & 1.9/4 \\
NGC 2110 & 99.1/100 & 2.6/3 & 4.7/4 & \textbf{9.8/3} & \textbf{9.9/4}\\
MCG+8-11-11 & 94.6/99 & 7.6/3 & \textbf{11.4/4}  & \textbf{9.6/3}  & \textbf{13.7/4} \\
Mrk 6 & 97.6/92 & 7.0/3 & \textbf{8.7/4}  & \textbf{7.2/3} & \textbf{8.0/4} \\
Mrk 110 & 116.0/100 & 3.7/3 & \textbf{12.1/4} & \textbf{14.7/3} & \textbf{20.5/4} \\
HE 1143-1800 & 94.0/99 & 3.0/3 & 5.1/4  & \textbf{8.4/3}  & \textbf{12.8/4} \\
IC 4329A(1) & 76.2/103 & 1.6/3 & 1.8/4  & 1.2/3  & 1.3/4 \\
NGC 5506(1) & 91.3/99 & 7.1/3 & 4.2/4 & \textbf{6.6/3}  & \textbf{8.0/4} \\
NGC 5548(1) & 84.1/100 & 0.2/3 & 0.0/4  & 0.1/3  & 1.7/4 \\
NGC 7213 & 111.7/99 & 1.1/3 & \textbf{7.4/4} & 0.7/3 & 0.7/4 \\
NGC 7469(1) & 122.0/100 & 8.2/3 & \textbf{13.6/4}  & \textbf{13.2/3} & \textbf{14.9/4} \\

\hline

\end{tabular}
\end{center}

\end{table}

\begin{table*}
\centering

\caption{The top panel displays the best-fit inclinations ($i$) and disc reflection strengths ($R$) for Models N07, ION, SRC and SRI. The lower panel shows the best-fit disc ionisation parameters ($\xi$) for Models ION and SRI and the best-fit emissivity indices ($q$) for Models SRC and SRI. Errors are given at a confidence level of 90\%.}
\label{tab:par}

\renewcommand{\arraystretch}{1.4}

\begin{center}
\begin{tabular}{lcccccccc}
\hline

Name & \multicolumn{4}{c}{$i$ ($^{\circ}$)} & \multicolumn{4}{c}{$R$} \\

& N07 & ION  & SRC & SRI & N07 & ION & SRC & SRI \\

\hline
Mrk 590 & $46^{+39 }_{-46}$ & $47^{+38}_{-47}$ & $48^{+37}_{-48}$ & $48^{+37}_{-48}$ & $0.41^{+0.58}_{-0.41}$ & $0.48^{+0.22}_{-0.21}$ & $0.43^{+0.91}_{-0.43}$ & $0.69^{+0.24}_{-0.23}$\\

NGC 2110  & $0^{+85}_{-0}$ & $0^{+51}_{-0}$ &  $61^{+2}_{-14}$ & $62^{+2}_{-11}$ & $0.14^{+0.34}_{-0.14}$ & $0.53^{+0.20}_{-0.41}$ &   $8.24^{+18.3}_{-6.08}$ & $5.02^{+4.98 }_{-0.30}$\\

MCG+8-11-11  & $85^{+0}_{-60}$ & $45^{+40}_{-8}$ & $73^{+6}_{-11}$ & $73^{+3}_{-11}$  & $3.18^{+4.04}_{-2.84}$ & $0.60^{+0.24}_{-0.23}$ &  $2.94^{+3.64}_{-2.26}$  & $2.67^{+0.25}_{-0.24}$\\

Mrk 6  & $26^{+59}_{-26}$ & $26^{+59}_{-7}$ &  $29^{+21}_{-10}$ & $30^{+14}_{-9}$  & $0.36^{+0.45}_{-0.30}$ & $0.55^{+0.37}_{-0.36}$ &  $0.88^{+5.43}_{-0.58}$ & $1.20^{+0.63}_{-0.59}$\\

Mrk 110 & $85^{+0}_{-85}$ & $36^{+19}_{-12}$ &  $60^{+7}_{-4}$ & $64^{+1}_{-6}$ & $0.80^{+1.02}_{-0.80}$ & $0.67^{+0.18}_{-0.18}$ &  $3.73^{+4.10}_{-0.68}$ & $5.01^{+0.08 }_{-0.20}$\\

HE 1143-1800 & $0^{+85}_{-0}$ & $0^{+85}_{-0}$   & $58^{+7}_{-58}$ & $58^{+12}_{-19}$ & $0.19^{+0.16}_{-0.19}$ & $0.46^{+0.20}_{-0.20}$ &  $6.40^{+1.85}_{-5.63}$ & $2.97^{+0.35}_{-0.33}$ \\

IC 4329A(1)  & $61^{+25}_{-60}$ & $62^{+23}_{-62}$ & $42^{+35}_{-42}$ & $85^{+0}_{-85}$ & $0.16^{+0.98}_{-0.16}$ & $0.18^{+0.21}_{-0.13}$ & $0.28^{+2.28}_{-0.22}$ & $0.50^{+0.27}_{-0.30}$ \\

NGC 5506(1)  & $46^{+4}_{-5}$ & $53^{+32}_{-53}$ &  $46^{+4}_{-4}$ & $46^{+5}_{-4}$  & $0.37^{+0.36}_{-0.22}$ & $0.13^{+0.06}_{-0.08}$  & $0.72^{+0.50}_{-0.44}$ & $0.69^{+0.33}_{-0.43}$ \\

NGC 5548(1)  & $15^{+75}_{-15}$ & $3^{+82}_{-3}$ & $64^{+21}_{-64}$ & $4^{+65}_{-4}$ & $0.06^{+0.33}_{-0.06}$ & $0.03^{+0.18}_{-0.03}$ &  $0.10^{+5.65}_{-0.10}$ & $0.77^{+0.31}_{-0.77}$ \\

NGC 7213  & $68^{+17}_{-68}$ & $0^{+64}_{-0}$ & $71^{+14}_{-71}$ & $69^{+16}_{-69}$ & $0.16^{+1.49}_{-0.16}$ & $0.11^{+0.07}_{-0.07}$ &  $1.57^{+2.71}_{-1.57}$ & $0.11^{+0.13}_{-0.11}$ \\

NGC 7469(1)  & $85^{+0}_{-53}$ & $85^{+0}_{-49}$  & $81^{+4}_{-6}$ & $76^{+7}_{-23}$ & $3.69^{+2.15}_{-1.84}$ & $0.97^{+0.29}_{-0.28}$ &  $3.57^{+2.44}_{-2.04}$ &  $1.30^{+0.31}_{-0.30}$\\

\hline
\end{tabular}

\begin{tabular}{lcccc}
\hline

Name & \multicolumn{2}{c}{$\xi$ (erg cm s$^{-1}$)} & \multicolumn{2}{c}{\textit{q}}\\
& ION & SRI & SRC & SRI\\
\hline

Mrk 590 &  $10^{+2860}_{-0}$ & $10^{+864}_{-0}$  & $3.0^{+6.0}_{-0.0}$ & $3.2^{+5.8}_{-0.2}$\\

NGC 2110 & $200^{+871}_{-190}$ & $19^{+669}_{-9}$  & $8.9^{+0.1}_{-2.6}$ & $9.0^{+0.0}_{-4.7}$\\

MCG+8-11-11 &  $200^{+1231}_{-190}$ & $29^{+200}_{-19}$   & $7.3^{+1.7}_{-2.6}$ & $8.8^{+0.2}_{-5.8}$\\

Mrk 6 &   $12^{+127}_{-2}$ & $11^{+96}_{-1}$  & $3.0^{+2.3}_{-0.0}$ & $3.0^{+1.8}_{-0.0}$\\

Mrk 110 &  $99^{+113}_{-89}$ & $54^{+47}_{-18}$  & $7.4^{+1.3}_{-2.2}$ & $8.9^{+0.1}_{-3.5}$\\

HE 1143-1800 &  $100^{+585}_{-90}$ & $77^{+116}_{-54}$  & $8.6^{+0.4}_{-5.0}$ & $6.9^{+2.1}_{-3.1}$\\

IC 4329A(1) & $21^{+165}_{-11}$ & $221^{+905}_{-210}$   & $3.0^{+6.0}_{-0.0}$ & $3.0^{+6.0}_{-0.0}$\\

NGC 5506(1) &  $555^{+300}_{-545}$ & $10^{+15}_{-0}$  & $3.0^{+1.0}_{-0.0}$ & $3.0^{+0.8}_{-0.0}$\\

NGC 5548(1) &  $5729^{+4271}_{-5719}$ & $4437^{+3979}_{-4427}$  & $3.0^{+6.0}_{-0.0}$ & $3.5^{+5.2}_{-0.5}$\\

NGC 7213 &  $498^{+637}_{-279}$ & $1341^{+8658}_{-1331}$  & $9.0^{+0.0}_{-6.0}$ & $9.0^{+0.0}_{-6.0}$\\

NGC 7469(1) &  $105^{+472}_{-82}$ & $100^{+142}_{-90}$  & $8.0^{+1.0}_{-5.0}$ & $7.1^{+1.9}_{-4.1}$\\

\hline
\end{tabular}

\end{center}
\end{table*}

\subsection{Strong relativistic effects -- Model SRC}

Our next set of models tested if the iron line is heavily blurred by concentrating the line emission in the central regions. This was accomplished firstly by modifying the emissivity profile to a single power-law with the index, \textit{q}, now a parameter free to vary between $\textrm{3} < q < \textrm{9}$. Secondly, we assumed a maximally spinning black hole by fixing the inner radius of the accretion disc to 1.24$\rg$.

We first assumed a cold accretion disc (PEXMON) to model the blurred disc reflection, with the ionised version investigated in the next section. This model is referred to as Model SRC (``Strong Relativistic effects with a Cold disc") and has the same XSPEC format as Model N07 (see Equation \ref{eq:xspec_n07}), although the emissivity parameters in the KDBLUR2 model component are different. As the break radius is irrelevant and the emissivity index is a free parameter, Model SRC has the same degrees of freedom as Model N07. For Model SRC, 95\% of the 10,000 Model NBR spectra are fit with $\dc < 6.1$ so this is used as our threshold to determine if Model SRC provides a good fit to the energy spectrum.

A significant detection of a blurred reflection component is made in 7/11 observations at a confidence level of 95\%: MCG+8-11-11, Mrk 6, Mrk 110, NGC 7469(1), NGC 2110, HE 1143-1800 and NGC 5506(1). The first four objects also gave a detection of an iron line with Model ION, which does not require a spinning black hole or a steep emissivity profile. Mrk 110 is the only one of these observations with a lower $\chi^{2}$ compared to Model ION, although Model SRC has one less degree of freedom. NGC 2110, HE 1143-1800 and NGC 5506(1) are the three observations that give a significant detection of the blurred reflection with Model SRC, but not with Model ION.

For the seven observations where a blurred reflection component is detected with Model SRC, the fit to the time-averaged spectrum is better than for Model N07. Furthermore, better constraints are generally obtained on the model parameters. The disc inclinations for MCG+8-11-11, Mrk 110 and NGC 7469(1) pegged at 85\deg\ with Model N07 while the lower limit of 0\deg\ was preferred for NGC 2110 and HE 1143-1800. The inclinations of these five observations no longer peg at the model boundaries as they range from $i = 58\de-81\de$. While the disc inclinations of Mrk 6 and NGC 5506(1) do not change from their Model N07 best-fit values, better constraints are placed on Mrk 6 with Model SRC. Apart from these two observations, the five other well-fit observations have disc inclinations that are much higher than those typical of the N07 broad line sources ($\langle i \rangle = 34.0\de \pm 6.4\de$), and hint at another reason why no broad line is clearly visible in these objects. The reflection strengths of these five observations are remarkably strong with $\sim R > \textrm{3}$. The magnitude of the blurring is revealed by the emissivity index. A steep emissivity profile with $q> \textrm{7.3}$ is found in the same five cases, which is indicative of very strong blurring. For NGC 2110, MCG+8-11-11, Mrk 110 and HE 1143-1800, the uncertainty of the emissivity index is not compatible with $q = 3$, which explains why reflection could not be detected with Model N07. The best-fit reflection strengths of Mrk 6 and NGC 5506(1) are smaller than the  five other well-fit observations, although the upper limit for Mrk 6 is consistent with being very strong. Furthermore, they require less concentrated emission from the central regions as their emissivity slopes peg at $q = 3$.

\subsection{Strong relativistic effects and ionisation of the accretion disc -- Model SRI}

In this Section, we test if a combination of strong relativistic effects and disc ionisation can reveal the blurred disc reflection in the narrow observations. This is achieved by replacing PEXMON in Model SRC with the ionised disc model REFLIONX. This model is termed Model SRI (``Strong Relativistic effects and an Ionised accretion disc"). Compared to Model N07, the black hole is maximally spinning, the accretion disc is ionised and the emissivity slope can be steeper than $q = 3$. A fit of Model SRI to the energy spectra is deemed acceptable at 95\% confidence level if $\dc$ relative to Model NBR is less than 6.7.

Model SRI provides a significant detection of the blurred reflection in the same 7/11 observations that are well-fit with Model SRC. Overall, Model SRI performs better than Model SRC for all seven observations but with one more degree of freedom, Model SRI outperforms Model SRC by $\dc > 3$ in only 3 observations: MCG+8-11-11, Mrk 110 and HE 1143-1800. This suggests that strong relativistic effects are the dominant factor in producing a good fit in the 4 other observations well-fit by Models SRC and SRI. This is in agreement with the low disc ionisation parameters of $\xi < 100$~erg cm s$^{-1}$ across the seven observations. Model SRI is also superior to Model ION for 5/7 observations. The exceptions are Mrk 6 and NGC 7213, with the former marginally preferring Model ION over Model SRI. As with Model SRC, a steep emissivity index of $q>$ 6.9 is preferred in 5/7 observations apart from Mrk 6 and NGC 5506(1), where it pegged at $q= 3$. Likewise, the disc inclinations found with Model SRI are very close to those given by Model SRC. They are higher than $i = \textrm{58}\de$ for 5/7 of the well-fit observations, although Mrk 6 and NGC 5506(1) have lower inclinations, consistent with the 18 ``relativistic-line" observations classified by N07. Unlike Model SRC, the upper error on the inclination for NGC 7469(1) does not extend to the model limit but it is still the highest of all the narrow observations. Another difference with Model SRC is that the REFLIONX reflection strengths of the 7 well-fit observations span a smaller range of $\textrm{0.7} < R < \textrm{5.0}$, although in most cases, this is still stronger than expected in the standard model.

\subsubsection{Reflection strength}

The reflection strengths derived from Model SRI, which includes both strong relativistic effects and disc ionisation, can be compared to those of the 18 N07 observations with a relativistically broadened iron line. Model SRI is used as it generally gave the best fit to most of the observations and all 11 narrow observations are considered. Using the maximum likelihood method of \citet{ma88}, the narrow-line observations have a mean reflection strength of $\langle R \rangle=1.63\pm$0.45 with an intrinsic spread of $\sigma_{R}= 1.50 \pm$0.32. This is $\sim 2$ times higher than the relativistic-line observations, who showed an average $\langle R \rangle=0.78\pm$0.13 with an intrinsic spread of $\sigma_{R}= 0.20 \pm$0.12. Thus, on the face of it (and somewhat paradoxically), the observations with no prominent broad line seem to have {\it stronger} reflection than those in which the broad, skewed line can be seen. 

Before drawing this conclusion, however, we consider the fact that the narrow-line observations also exhibit on average a higher disc inclination, $\langle i \rangle = 62.9\de \pm 7.7\de$ with Model SRI, compared to $\langle i \rangle = 34.0\de \pm 6.4\de$ for the relativistic observations. This in itself provides a partial explanation for the difficulty in detecting the broad lines in N07, as the higher inclination sources will suffer stronger Doppler effects, further blending the line into the continuum. The fact that the higher inclination sources also show larger reflection fractions is somewhat suspicious, as it implies the total strength of the reflection is similar, whereas it should be weaker in the case of the highly inclined sources. To compare these more directly, we use a more robust measure of the reflection strength, which we call the reflected flux fraction (RFF). This is the ratio of the blurred reflected flux to the continuum flux between 2.5 -- 10~keV. The mean RFF of the narrow-line observations is 0.17$\pm$0.07 with an intrinsic dispersion of 0.11$\pm$0.03. This is only slightly greater than the mean RFF of the 18 N07 relativistic-line objects, 0.13$\pm$0.03 with an intrinsic dispersion of 0.03$\pm$0.01. Therefore, after accounting for inclination effects, the strength of the blurred disc reflection in the narrow-line observations is similar to that of the observations with a relativistically broadened iron line.

\section{Discussion}

We have reanalysed the \xmm\ spectra of 11 Seyfert galaxies classified by N07 as not requiring a broad, accretion disc iron line in a simple analysis. By fitting models which allow the disc to be ionised, or to have stronger relativistic effects in a Kerr geometry, we have found that 8 of these 11 observations, in fact, do show evidence for a significant blurred reflection component. The mystery regarding the lack of apparent relativistic reflection in these sources in the N07 analysis can therefore be explained. The reason is that the most obvious feature in the reflection spectrum --- the broad iron K${\alpha}$ line --- is rendered indistinguishable from the underlying continuum, by a combination of blending and Comptonisation in an ionised disc, strong relativistic effects and, in some cases, a high disc inclination.

Although the models and parameters are rather degenerate and somewhat difficult to disentangle, the most important effect seems to be that of strong relativistic blurring. A mildly ionised reflection component can be detected in MCG+8-11-11, Mrk 6, Mrk 110, NGC 7213 and NGC 7469(1) without the need for a steep emissivity power-law or a spinning black hole. On the other hand, seven objects can be well-fit with a cold disc as long as the emission can be concentrated in the very innermost regions where the Kerr metric holds sway. In reality, it seems that both effects are likely at play in many cases. For example, the strongest detection of blurred reflection is made in Mrk 110 with a combination of both ionisation and strong gravity.

The blurred disc reflection in the narrow-line observations is as strong, if not slightly stronger, as it is in the N07 observations that are well-fit with a relativistically broadened iron line. Furthermore, a steep emissivity profile is found in at least five of the observations that give a significant detection of the reflection spectrum. A combination of strongly blurred reflection and a centrally concentrated disc emissivity is consistent with gravitational light bending. This predicts that if the X-ray source is close to the plane of the disc, continuum photons will be bent towards the accretion disc and black hole instead of escaping to infinity. As a result, the intrinsic continuum is suppressed and the anisotropic illumination of the disc can produce highly smeared reflection. The strength of the reflection can also vary much more from source-to-source than in a simple Newtonian geometry, because it depends on the degree of light bending and relativistic focusing. The line of sight to the discs with high inclinations could be obscured by the molecular torus, but a misalignment of the disc with the torus due to e.g.\ chaotic accretion \citep{ki06} can result in Seyfert 1s with high inclinations.

The heavily blurred iron lines are associated with strong Compton humps in the 7 narrow-line observations that are well-fit with Models SRC or SRI. Indeed in 4/7 of the observations, the reflection hump is stronger than the continuum above 10~keV. While the bandpass of \xmm\ is limited to 12~keV, \suz\ is suited to observing the high energy X-ray spectra above 10~keV. The only two narrow-line objects that have been observed with \suz\ are NGC 7469 \citep{pa11} and NGC 2110 \citep{re06}. The hard X-ray fluxes in both are consistent with the predictions of Model SRI using the parameters listed in Table \ref{tab:par}.

\subsection{Increasing the signal-to-noise ratio}

Part of the difficulty in identifying the blurred disc reflection in some of the narrow-line observations is their short observation length i.e.\ the low signal-to-noise ratio. It is possible that longer observations of the same objects may reveal the disc reflection. To test this, we took the Model SRI best-fit spectra for the 7 observations that are well-fit with this model and created fake spectra with exposure times of 100~ks. Fitting the fake spectra with Model ION gave a good fit to Mrk 6, NGC 5506(1) and NGC 7469(1) at 99\% confidence level. Therefore, in these three observations, blurred disc reflection can be detected without the requirement of a spinning black hole and emissivity slope greater than $q = 3$. However, in the 4 remaining observations, Model ION does not give a good fit to spectra with exposure times as large as 500~ks. This suggests that the iron line will not be resolved in some cases even if the S/N is extremely good unless extreme blurring effects are included in models.

\subsection{Missing blurred reflection}

Incorporating our results with those of N07, a relativistically or non-relativistically broadened reflection component is now detected in 34/37 observations of the total sample. The three narrow observations that do not show evidence for a broad iron line component with the ionised and/or blurred iron line models are: Mrk 590, IC 4329A(1) and NGC 5548(1). A high ionisation parameter of $\xi = 4437$~erg cm s$^{-1}$ with Model SRI could explain why no disc reflection is detected in NGC 5548(1). Mrk 590 and IC 4329A(1) are similar to Mrk 6 and NGC 5506(1) in that they are best-fit with a small reflection strength and an emissivity index close to $q=3$. However, while Mrk 590 has a moderate disc inclination, our view of the disc in IC 4329A(1) is edge-on in Model SRI. In this case, if it is true that the disc inclination is high but the relativistic effects are not strong, the reflection and iron line would be expected to be faint and extremely difficult to identify in \xmm\ spectra.

Overall, our results provide a solution to the long-standing problem of why some sources do not exhibit broad iron lines. Such objects have posed a difficulty for models of AGN as a relativistically broadened iron line and reflection component should be a natural outcome of matter accreting onto a supermassive black hole via an accretion disc that is illuminated by hard X-rays. We have tested if the reflection apparently appears to be ``missing" because the accretion disc is ionised and/or the reflection is heavily smeared. The latter effect can be implemented by both a steep emissivity profile and a spinning black hole. The combination of these two factors can account for the missing blurred reflection in 8/11 sources although all 11 observations can allow for this type of reflection. We find that a steep emissivity profile is essential in smearing the reflection to such an extent that it is difficult to separate from the continuum.

Furthermore, the stronger Doppler effects associated with a high disc inclination around $60\de$ enhances the blending of the line into the continuum. However, we are still able to identify this spectral component as the observations generally have reflection strengths that are larger than expected in a geometry where the disc is illuminated from above by a point source. The strong and heavily smeared reflection is consistent with the expectation of the gravitational light bending models.

\section*{Acknowledgements}

This paper is based on observations taken with the XMM-Newton Satellite, an ESA science mission with contributions by ESA Member States and USA. KN thanks the Royal society for support and SB acknowledges funding from a Science and Technology Facilities Council (STFC) studentship.

\bsp

\label{lastpage}

\end{document}